\documentclass[
    ,final            
  ]
  {aipproc}

\layoutstyle{8x11double}

\def\lsim{\raise0.3ex\hbox{$\;<$\kern-0.75em\raise-1.1ex\hbox{$\sim\;$}}}
\def\gsim{\raise0.3ex\hbox{$\;>$\kern-0.75em\raise-1.1ex\hbox{$\sim\;$}}}

\def\bsg{$b\to s\gamma$}

\def\asusy{a^{\rm SUSY}_\mu}

\DeclareMathAlphabet   {\mathsc}{OT1}{cmr}{m}{sc} 
 
\def\[{\left [} 
\def\]{\right ]} 
\def\({\left (} 
\def\){\right )}

\newcommand{\gappeq}{\mathrel{\rlap {\raise.5ex\hbox{$>$}} 
{\lower.5ex\hbox{$\sim$}}}} 
\newcommand{\lappeq}{\mathrel{\rlap{\raise.5ex\hbox{$<$}} 
{\lower.5ex\hbox{$\sim$}}}}

\newcommand{\bea}{\begin{eqnarray}}
\newcommand{\eea}{\end{eqnarray}}

\begin{document}

\title{Gamma-Ray Excess from the Galactic Center and Supergravity Models}


\author{\underline{Y. Mambrini}}{
  address={Laboratoire de Physique Th\'eorique, Universit\'e Paris--Sud,
F--91405 Orsay, France}
}

\author{C. Mu\~noz}{
  address={Departamento de F\'isica Te\'orica C-XI, Universidad Aut\'onoma
de Madrid, Cantoblanco, 28049 Madrid, Spain}
}

\author{E. Nezri}{
  address={Service de Physique Th\'eorique, CP225, Universit\'e Libre de
Bruxelles, 1050 Bruxelles, Belgium}
}

\author{F. Prada}{
  address={Instituto de Astrof\'isica de Andaluc\'ia (CSIC), E-18008 Granada, Spain}
}

\begin{abstract}

We analyse the effect of the compression of the dark matter
due to the infall of baryons to the galactic center on the
gamma-ray flux. In addition, we also consider the effect of non-universal 
supersymmetric soft terms. This analysis shows that neutralino dark matter 
annihilation can give rise to signals largely reachable by
future experiments like GLAST. This is a remarkable result if we 
realise that direct detection experiments will only be able to cover 
a small region of the parameter space.
Actually, in this SUGRA framework we have also been able to  
fit present excess from EGRET and CANGAROO using different
non-universal scenarios,
and even fit the data from both experiments with only one scenario.
We have also studied the recent HESS data implying a neutralino
heavier than 12 TeV. Because of such a heavy neutralino,
it is not natural to find solutions in the SUGRA framework.
Nevertheless we have carried out a quite model-independent
analysis, and found the conditions required on the particle physics
side to fit the HESS data thanks to dark matter annihilation.

\end{abstract}

\maketitle

\section{Introduction}

It is now well established that luminous matter makes up only a small
fraction of the mass observed in Universe. 
A weakly interacting massive particle (WIMP) is one of the leading
candidates for the ``dark'' component of the Universe.
One of the most promising methods for the
indirect detection of WIMPs consists of detecting the gamma rays produced by
their annihilations in the galactic halo. 
Concerning the nature of WIMPs, 
the best motivated candidate is the lightest
neutralino, a particle predicted by the 
supersymmetric (SUSY) extension of the standard model \cite{mireview}. 
We implement in our analysis \cite{Mambrini:2005vk} the lower bounds on the 
masses of SUSY particles and Higgs boson, as well as the experimental
bounds on the branching ratio of the \bsg\ process and on 
$\asusy$, for which the more stringent constraint from $e^+e^-$ 
disfavors 
important regions of the SUGRA parameter space 
(see e.g. Ref. \cite{Bertone:2004ag,Mambrini:2004ke}). 
In addition,
we have also taken into account  the last data concerning the 
$B_s \to \mu^+ \mu^-$ branching ratio.

\section{The gamma-ray flux}

For the continuum of gamma rays, the observed differential flux at the Earth  
coming from a direction forming an angle
$\psi$ with respect to the galactic center is
\begin{equation}
\Phi_{\gamma}(E_{\gamma}, \psi)
=\sum_i 
\frac{dN_{\gamma}^i}{dE_{\gamma}}
 \langle\sigma_i v\rangle \frac{1}{8 \pi m_{\chi}^2}\int_{line\ of\ sight} \rho^2
\ dl\ ,
\label{Eq:flux}
\end{equation}
\noindent 
where the discrete sum is over all dark matter annihilation
channels,
$dN_{\gamma}^i/dE_{\gamma}$ is the differential gamma-ray yield,
$\langle\sigma_i v\rangle$ is the annihilation cross section averaged over its
velocity
distribution, $m_{\chi}$ is the mass of the dark matter particle,
and $\rho$ is the dark matter density. 
Actually, when comparing to experimental data
one must consider the integral of $\Phi_{\gamma}$
over the spherical region of solid angle 
$\Delta \Omega$ given by the angular acceptance of the detector
which is pointing towards the galactic center. 
For example, for EGRET $\Delta \Omega$  is about
$10^{-3}$ sr whereas for GLAST, CANGAROO, and HESS it is $10^{-5}$ sr.
Note that neutralinos move at galactic velocity and therefore
their annihilation occurs at rest.

\section{The adiabatic compression model}

Recently, SUSY dark matter candidates have been studied
in the context of realistic halo models including
baryonic matter \cite{Prada:2004pi}.
Indeed, since the total mass of the inner galaxy is dominated by baryons,
the dark matter distribution is likely to have been influenced by 
the baryonic potential. In particular, its density is increased, and 
as a consequence typical halo profiles such as
Navarro, Frenk and White (NFW) \cite{Navarro:1996he} and
Moore et al. \cite{Moore:1999gc} have
a more singular behaviour near the galactic center.
The conclusion of the work in Ref.~\cite{Prada:2004pi} 
is that the gamma-ray flux produced by the annihilation of neutralinos
in the galactic center is increased significantly,
and is within the sensitivity of incoming experiments, 
when density profiles with baryonic compression are taken into account.
Indeed, highly cusped profiles are deduced
from N-body simulations. In particular, NFW \cite{Navarro:1996he} 
obtained a profile with a behaviour $\rho(r)\propto r^{-1}$ at small
distances. A more singular behaviour,  $\rho(r)\propto r^{-1.5}$, was 
obtained by Moore et al. \cite{Moore:1999gc}. 
However, these predictions are valid only for halos 
without baryons. One can improve simulations in a more realistic way by taking into
account the effect of the normal gas (baryons). This loses its 
energy through radiative
processes falling to the central region of forming galaxy.
As a consequence of this redistribution of mass,
the resulting gravitational potential is deeper, and the dark matter
must move closer to the center increasing its density. 

This increase in the dark matter density is often treated using
adiabatic invariants. The present form of the adiabatic compression model
was numerically and analytically studied by Blumental et
al. \cite{Blumenthal:1985qy}. 
This model assumes spherical symmetry, circular orbit for
the particles, and conservation of the angular momentum $M(r) r = $ const., where
$M(r)$ is the total mass enclosed within radius $r$. The mass distributions in the initial 
and final configurations are therefore related by 
$M_i(r_i) r_i = [M_b(r_f) + M_{DM}(r_f)] r_f$,
where $M_i(r)$, $M_b(r)$ and $M_{DM}(r)$ are the mass profile
of the galactic halo before the cooling of the baryons (obtained through N-body simulations),
the baryonic composition of the Milky Way observed now, and the to be determined
dark matter component of the halo today, respectively. 
This approximation was tested in numerical simulations \cite{Gnedin}.
Nevertheless, a more precise approximation can be obtained including
the possibility of elongated orbits \cite{Prada:2004pi}. 
The models and constraints that we used in this work for the Milky Way can 
be found in Table I of Ref.~\cite{Prada:2004pi}.
As one can see in \cite{Mambrini:2005vk}, at small $r$ the dark matter density profile following the adiabatic cooling
of the baryonic fraction is a steep power law $\rho \propto r^{-\gamma_c}$
with $\gamma_c \approx 1.45 (1.65)$ for a $\rm{NFW_c}$($\rm{Moore_c}$)
compressed model.

\section{Supersymmetric Models}

As discussed in detail in 
Ref.~\cite{Mambrini:2004ke} in the context of indirect detection,
$\sigma_i$ can be increased 
in different
ways when the structure of mSUGRA for the soft terms is abandoned. 
In particular, it is possible to enhance the annihilation
channels involving exchange of the CP-odd Higgs, $A$, by reducing the
Higgs mass. In addition, it is also possible to  increase the Higgsino 
components of the lightest neutralino. 
Thus annihilation channels through Higgs exchange become more important
than in mSUGRA.
This is also the case for $Z^-$, $\chi_1^\pm$, and  $\tilde{\chi}_1^0$-exchange
channels.
As a consequence, the gamma-ray flux will be increased.

In particular, the most important effects are produced
by the non-universality of Higgs and gaugino masses.
These can be parameterised, at $M_{GUT}$, as follows
\begin{equation}
  m_{H_{d}}^2=m^{2}(1+\delta_{1})\ , \quad m_{H_{u}}^{2}=m^{2}
  (1+ \delta_{2})\ ,
  \label{Higgsespara}
\end{equation}
and 
\begin{eqnarray}
  M_1=M\ , \quad M_2=M(1+ \delta'_{2})\ ,
  \quad M_3=M(1+ \delta'_{3})
  \ ,
  \label{gauginospara}
\end{eqnarray}
We concentrated in \cite{Mambrini:2005vk} 
our analysis on the following representative cases:
\begin{eqnarray}
a)\,\, \delta_{1}&=&0\ \,\,\,\,\,\,\,\,,\,\,\,\, \delta_2\ =\ 0
\,\,\,\,\,\,\,\,\,,\,\,\,\, 
\delta'_{2,3}\ =\ 0\ , 
\nonumber\\
b)\,\, \delta_{1}&=&0\ \,\,\,\,\,\,\,\,,\,\,\,\, \delta_2\ =\ 1\
\,\,\,\,\,\,\,\,,\,\,\,\, 
\delta'_{2,3}\ =\ 0\ , 
\nonumber\\
c)\,\, \delta_{1}&=&-1\ \,\,\,\,,\,\,\,\, \delta_2\ =\ 0\
\,\,\,\,\,\,\,\,,\,\,\,\, 
\delta'_{2,3}\ =\ 0\ , 
\nonumber\\
d)\,\, \delta_{1}&=&-1\ \,\,\,\,,\,\,\,\, \delta_2\ =\ 1\
\,\,\,\,\,\,\,\,,\,\,\,\, 
\delta'_{2,3}\ =\ 0\ \,\, , 
\nonumber\\
~~e)\,\, \delta_{1,2}&=&0\ \,\,\,\,\,\,\,\,,\,\,\,\, \delta'_{2}\ =\ 0
\,\,\,\,\,\,\,\,\,,\,\,\,\, 
\delta'_{3}\ =\ -0.5\ ,
\nonumber\\
~~f)\,\, \delta_{1,2}&=&0\ \,\,\,\,\,\,\,\,,\,\,\,\, \delta'_{2}\ =\ -0.5
\,\,\,\,\,\,\,\,\,,\,\,\,\, 
\delta'_{3}\ =\ 0\ .
\label{3cases}
\end{eqnarray}
\noindent Case {\it a)} corresponds to mSUGRA with universal soft terms,
cases {\it b)}, {\it c)} and  {\it d)} 
correspond to  non-universal Higgs masses, 
and finally cases {\it e)} and {\it f)} to non-universal gaugino masses.
The cases {\it b)},  {\it c)},  {\it d)}, and {\it e)} were discussed in Ref.~\cite
{Mambrini:2004ke},
and are known to produce gamma-ray fluxes larger than in mSUGRA, whereas case
{\it f)} will be of interest when discussing heavy WIMP signals predictions 
in the perspective of atmospheric Cherenkov telescopes like e.g. CANGAROO.

\section{Confronting experiments}

\subsection{EGRET}

\begin{figure}
  \includegraphics[height=.2\textheight]{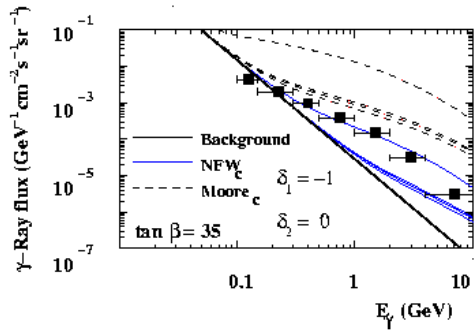}
  \caption{Gamma--ray spectra $\Phi_{\gamma}(E_{\gamma})$
from the galactic center as functions of the photon energy
for the SUGRA case {\bf c)} discussed in Eq.~(\ref{3cases})
for  $\tan \beta=35$, $A=0$ and $\mu > 0$, compared with data from the 
EGRET experiment. NFW and Moore et al. profiles with
adiabatic compression are used with $\Delta \Omega \sim 10^{-3}$ sr.
All points shown after a scan on $m$ and $M$ (0 -- 2000 GeV) 
fulfil the accelerator constraints discussed in the text, 
and WMAP bounds.}
\label{fig:EGRET}
\end{figure}

The EGRET telescope on board of the Compton Gamma-Ray Observatory 
has carried out the first all-sky survey in high-energy
gamma-rays ($\approx$ 30 MeV -- 30 GeV) over a period of 5 years,
from April 1991 until September 1996. As a result of this survey, it has 
detected a signal \cite{EGRET} 
above about 1 GeV, with a value for the flux of about
$10^{-8}$ cm$^{-2}$\ s$^{-1}$, that apparently cannot be explained with the 
usual gamma-ray background. 
The source, possibly diffuse rather than pointlike, is located within 
the $1.5^o$ ($\Delta \Omega \sim 10^{-3}$ sr) of the galactic center.
Due to the lack of precision data in the high energy bins, it seems impossible however to distinguish any annihilation channel leading to this photon excess. 
The results can be seen in Fig.~\ref{fig:EGRET},
where case {\bf c)} of Eq.~(\ref{3cases}) have been studied
for a scan on $m$ and $M$ from 0 to 2 TeV and tan$\beta=35$.
Let us remark that it is possible to differentiate each point of the parameter space ($m$, $M$) by its gamma-ray spectrum. The higher fluxes for instance
corresponds to the closing of the A-pole, 
whereas the lower flux spectrum is obtained through the opening of
the A-pole (see e.g. point {\bf B} of Fig. 3 in \cite{Mambrini:2005vk}).

\subsection{CANGAROO-GLAST}

Recently,  the CANGAROO--II atmospheric Cherenkov telescope has made a
significant detection of gamma rays from the Galactic center 
region \cite{Cangaroo1}. In particular, 
the collaboration has published the spectrum obtained in six energy bins, 
from 200 GeV to 3 TeV.
Observations taken during 2001 and 2002 have detected a statistically
significant excess at energies greater than 
$\sim 250$ GeV, with an integrated flux of 
$\sim 2 \times 10^{-10} ~ \mathrm{photons}~ \mathrm{cm}^{-2}~ \mathrm{s}^{-1}$.
These measurements indicate a very soft spectrum $\propto E^{-4.6 \pm 0.5}$.

 It is interesting to see whether it is possible
to obtain such a candidate in SUGRA scenarios imposing the accelerator
and WMAP constraints, and withing the framework of adiabatically
compressed halos.
The baryonic cooling effect on the fluxes gives us the order of
magnitude needed 
to fit with both data with a 1 TeV neutralino in the 
non--universal case {\it e)} with $M_3=0.5 M$.
It is worth noticing that the CANGAROO-II collaboration in \cite{Cangaroo1}
pointed out already that the EGRET and 
CANGAROO-II data can be relatively smoothly 
connected with a cutoff energy of 1--3 TeV. 
Typical points
of the parameter space fullfilling all experimental constraints and fitting
both set of data lie between ($m=800$ GeV, $M= 800$ GeV) and
($m=3$ TeV, $M=3$ TeV).

\begin{figure}
  \includegraphics[height=.2\textheight]{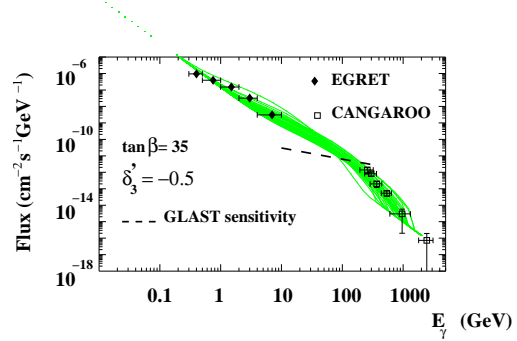}
  \caption{Gamma-ray spectra from the galactic center as functions 
of the photon energy for the non--universal case {\it e)} with $M_3=0.5 M$,
compared with data for EGRET and CANGAROO--II experiments and
the expected GLAST sensitivity. Here only the average profile
defined in Sect.~\ref{crucial} with adiabatic compression, 
$\rm{NFW'_c}$,
is used.  }
\label{fig:EGRETCANGAROO}
\end{figure}

It is also interesting to see the complementarity
of GLAST with EGRET 
and CANGAROO. GLAST will perform an all-sky survey detection of fluxes 
with energy from 1 GeV to 300 GeV, 
exactly filling the actual lack of experimental data 
in this energy range (see Fig. \ref{fig:EGRETCANGAROO}),
and checking the CANGAROO results. Indeed, we have calculated
that the integrated gamma ray flux for such a signal will be around 
$5 \times 10^{-11} ~ 
\mathrm{cm^{-2}} ~ \mathrm{s^{-1}}$. 
We have shown this sensitivity curve in Fig. \ref{fig:EGRETCANGAROO} 
for  $\Delta \Omega = 10^{-5}$, which is the typical detector acceptance, following
the prescriptions given in \cite{Fornengo}. We clearly see that GLAST 
will finish to cover
the entire spectrum.

\subsection{HESS}

The HESS Cherenkov telescope experiment has recently published new
data on gamma rays, detecting a signal from the Galactic Center
\cite{HESS}. The measured flux and spectrum differ substantially from
previous results, in particular those reported by the CANGAROO
collaboration, exhibiting a much harder power--law energy spectrum with 
spectral index of about $-2.2$ and extended up to 9 TeV.
The authors of \cite{HESS} already pointed out that if we assume that
the observed gamma rays represent a continuum annihilation spectrum, 
we expect $m_{\chi} \gsim 12$ TeV. Actually such a heavy neutralino-LSP is 
not natural in the framework of a consistent supergravity model when we 
impose the renormalisation group equations and radiative electroweak symmetry
breaking.

Although in \cite{Mambrini:2005vk} 
we performed some scans in all non universality directions using 
numerical dichotomy methods, no point in the parameter space in any 
non-universal case studied was able to give a several 10 TeV 
neutralino satisfying WMAP constraint but this can be sensitive to the RGE 
and relic density calculation codes. 

\begin{figure}
  \includegraphics[height=.2\textheight]{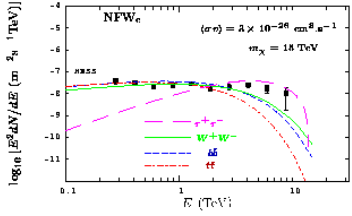}
  \caption{Dark matter annihilation versus HESS data for NFW compressed 
halo models.}
	\label{fig:HESS}
\end{figure}

Nevertheless, without RGE and taking soft parameters at the electroweak scale,
the constraints are easier to evade. On top of that, in a very effective  
approach using  completely free parameters and couplings in cross sections, 
neutralinos with 
$m_{\chi}\sim 10$ TeV and $\Omega_{\chi} h^2 \sim WMAP$ may certainly be 
fine tuned. We did not adopt such approaches since the MSSM is motivated by 
high energy and theoretical considerations.
We analyzed  in a quite model-independant way the conditions 
required on the particle physics field to fit with the HESS data thanks 
to dark matter annihilation.
The results are shown in Fig. \ref{fig:HESS}. We point the
fact that we do not only
 compare the spectrum shape of the signal with possible dark matter 
annihilation explanation. Indeed, it should be noticed that our compressed
halo profiles give rise to absolute gamma fluxes within the HESS data order of
magnitude with $\langle \sigma v \rangle$ values in agreement with the WMAP
requirement.
It is worth noticing here than a similiar and more complete analysis
was done recently by S. Profumo in \cite{Profumo:2005xd}.

\section{Conclusion}

We have analysed the effect of the compression of the dark matter
due to the infall of baryons to the galactic center on the
gamma-ray flux. In addition, we have also consider the effect of non-universal 
soft terms, that arises naturally in string motivated framework
\cite{Mambrini}.
This analysis shows that neutralino dark matter 
annihilation can give rise to signals largely reachable by
future experiments like GLAST. 
This is a remarkable result if we realise that 
direct detection experiments will
only be able to cover a small region of the parameter space.
Actually, in this SUGRA framework we have also been able to  
fit present excess from EGRET and CANGAROO using different
non-universal scenarios,
and even fit the data from both experiments with only one scenario.
We have also studied the recent HESS data implying a neutralino
heavier than 12 TeV. 
We have carried out a quite model-independent
analysis, and found the conditions required on the particle physics
side to fit the HESS data thanks to dark matter annihilation.

In any case, we must keep in mind that the current data obtained by the 
different gamma-rays observations from the Galactic
Center region by these three experiments
do not allow us to conclude about a dark
matter annihilation origin rather than other, less exotic astrophysics sources. 
Fortunately, this situation may change with 
the improvement of angular 
resolution and energy sensitivity 
of future detectors like GLAST, which will be a complementary
source of gamma-ray data.

\begin{theacknowledgments}
Y.M. want to thank the organisation commity  for having
made him discover the true mystery of {\it Kimchis}. 
\end{theacknowledgments}


\end{document}